\documentclass[12pt, preprint]{aastex}
\begin{document}
\title{The 2004 Hyperflare from SGR 1806-20: Further Evidence for Global 
Torsional Vibrations} \author{Tod E. Strohmayer$^1$ and Anna
L. Watts$^2$} \altaffiltext{1}{Exploration of the Universe Div.,
NASA/GSFC, Greenbelt, MD 20771; stroh@milkyway.gsfc.nasa.gov}
\altaffiltext{2}{Max Planck Institut f\"ur Astrophysik,
Karl-Schwarzschild-Str. 1, 85741 Garching, Germany; anna@mpa-garching.mpg.de}
\begin{abstract}

We report an analysis of the archival Rossi X-ray Timing Explorer
(RXTE) data from the December 2004 hyperflare from SGR 1806-20.  In
addition to the $\approx 90$ Hz QPO first discovered by Israel et al.,
we report the detection of higher frequency oscillations at $\approx
150$, 625, and 1,840 Hz.  In addition, we also find indications of
oscillations at $\approx 720$, and 2,384 Hz, but with lower
significances. The 150 Hz QPO has a width (FWHM) of about 17 Hz, an
average amplitude (rms) of $6.8\%$, and is detected in average power
spectra centered on the rotational phase of the strongest peak in the
pulse profile. This is approximately half a cycle from the phase at
which the 90 Hz QPO is detected. The 625 Hz oscillation was detected
in an average power spectrum from nine successive cycles beginning
approximately 180 s after the initial hard spike. It has a width
(FWHM) of $\approx 2$ Hz and an average amplitude (rms) during this
interval of $8.5\%$. We find a strong detection of the 625 Hz
oscillation in a pair of successive rotation cycles begining about 230
s after the start of the flare.  In these cycles we also detect the
1,840 Hz QPO. When the 625 Hz QPO is detected we also confirm the
simultaneous presence of 30 and 92 Hz QPOs.  The centroid frequency of
the 625 Hz QPO detected with RXTE is within 1 Hz of the $\approx 626$
Hz oscillation recently found in RHESSI data from this hyperflare by
Watts \& Strohmayer.  We argue that these new findings provide further
evidence for a connection of these oscillations with global
oscillation modes of neutron stars, in particular, the high frequency
signals may represent toroidal modes with at least one radial node in
the crust. We discuss their implications in the context of this model,
in particular for the depth of neutron star crusts.

\end{abstract}

\section{Introduction}

Recent studies of high time resolution data from two magnetar
hyperflares recorded by the Proportional Counter Array (PCA) onboard
the Rossi X-ray Timing Explorer (RXTE) have resulted in the discovery
of a new phenomenon associated with these events. Both the December,
2004 event from SGR 1806-20, and the August, 1998 hyperflare from SGR
1900+14 produced fast, rotation-phase-dependent X-ray oscillations in
the 20 - 150 Hz range.  First, Israel et al. (2005) reported the
discovery of $\approx 18$, 30, and 90 Hz quasi-periodic oscillations
(QPO) in the December, 2004 event, and suggested that the 30 and 90 Hz
QPOs could be linked with seismic vibrations of the neutron star
crust. Strohmayer \& Watts (2005) then reported the discovery of a
sequence of QPOs in the SGR 1900+14 hyperflare.  They found a set of
frequencies; 28, 53.5, 84, and 155 Hz, which could plausibly be
associated with a sequence of low $l$ toroidal modes (denoted $_l
t_0$) of the elastic neutron star crust (see, for example, Hansen \&
Cioffi 1980; McDermott, van Horn \& Hansen 1988; Duncan 1998; Piro
2005).  In both hyperflares the oscillations are episodic, that is,
their amplitudes vary considerably with time and rotational phase.
Most recently, Watts \& Strohmayer (2006) examined Ramaty High Energy
Solar Spectroscopic Imager (RHESSI) data for the SGR 1806-20 event,
confirmed the presence of the 18 Hz and 90 Hz QPOs, and found evidence
for additional oscillations at 26 Hz and 626 Hz.

The similar phenomenology of the oscillations in the two sources, as
well as the closeness of some of the measured frequencies argues
rather convincingly that we are seeing the same physical process in
each case.  The connection with torsional modes of the crust seems
plausible for several reasons; 1) The observed frequencies are
consistent with theoretical expectations for such modes, and can be
more or less self-consistently associated with a sequence of modes
with varying spherical harmonic index, $l$. 2) The magnetic
instability which powers the hyperflares is likely associated with
large scale fracturing of the neutron star crust \citep{flo77, tho95,
dun98, td01, sch05}, and will almost certainly generate seismic
motions within the star. 3) The strong phase dependence argues for a
mechanism associated with particular sites on the stellar surface,
such as a fracture zone or magnetic field bundle. 4) Mechanical
motions provide a natural explanation for the relatively high
coherence of the oscillations.  5) Horizontal motions of the crust
could modulate the beaming pattern associated with the strong magnetic
field, providing a mechanism to modulate the X-ray flux. Moreover,
beaming can act as an ``amplifier,'' producing potentially large X-ray
modulations from modest horizontal displacements.  Although the
present evidence for torsional modes is very suggestive, it is not yet
definitive. Levin (2006), for example, argues that toroidal modes may
damp too quickly to account for the detection of oscillations some
minutes after the onset of the flare.  In addition some of the
detected frequencies (18 Hz, 26 Hz) do not fit easily into current
torsional mode models, without invoking magnetic splitting or other
complications.

In our previous study of the RXTE data from the SGR 1900+14 hyperflare
we computed average power spectra for different rotational phases and
time intervals during the event.  This method should be more sensitive
to oscillations which are localized in rotational phase, as borne out
in the SGR 1900+14 study.  In this paper we report the results of a
similar, phase averaged timing study on the now public archival RXTE
data from the SGR 1806-20 hyperflare.  Using this method we find
strong evidence for additional, higher frequency oscillations during
the flare.

\section{Observations and Data Analysis}

An overview of the RXTE data recorded from the December, 2004
hyperflare from SGR 1806-20 is given by Israel et al. (2005).  Data
were recorded in the ``Goodxenon\_2s'' mode that allows for time
resolution up to $\approx 1$ $\mu$s.  The hyperflare intensity
profile, folded at the rotational frequency (see Figure \ref{fig1}),
shows 3 peaks. In subsequent discussions we will refer to these as
peaks 1, 2, and 3, in order of decreasing peak intensity,
respectively. We will also refer to the region of pulse phase between
peaks 2 and 3 as the ``interpulse'' region.

\subsection{Kilohertz Oscillations}

Since previous studies have shown the $\approx$ 90 Hz QPO to be
extremely robust, and its approximate location in rotational phase has
been linked to the interpulse region, by both Israel et al. (2005) and
Watts \& Strohmayer (2006), we began our study by computing average
power spectra around this phase range, but for a sequence of different
time intervals during the flare.  We used 3 s intervals to compute
each individual power spectrum and we initially restricted our search
to a Nyquist frequency of 4096 Hz, although these PCA data can, in
principle, be used to sample to much higher frequencies.  We found a
significant signal at 625 Hz in power spectra computed from
approximately the last third of the hyperflare. We now discuss the
detection, significance and properties of this oscillation.

\begin{figure}
\begin{center}
\includegraphics[width=6in, height=6in, clip]{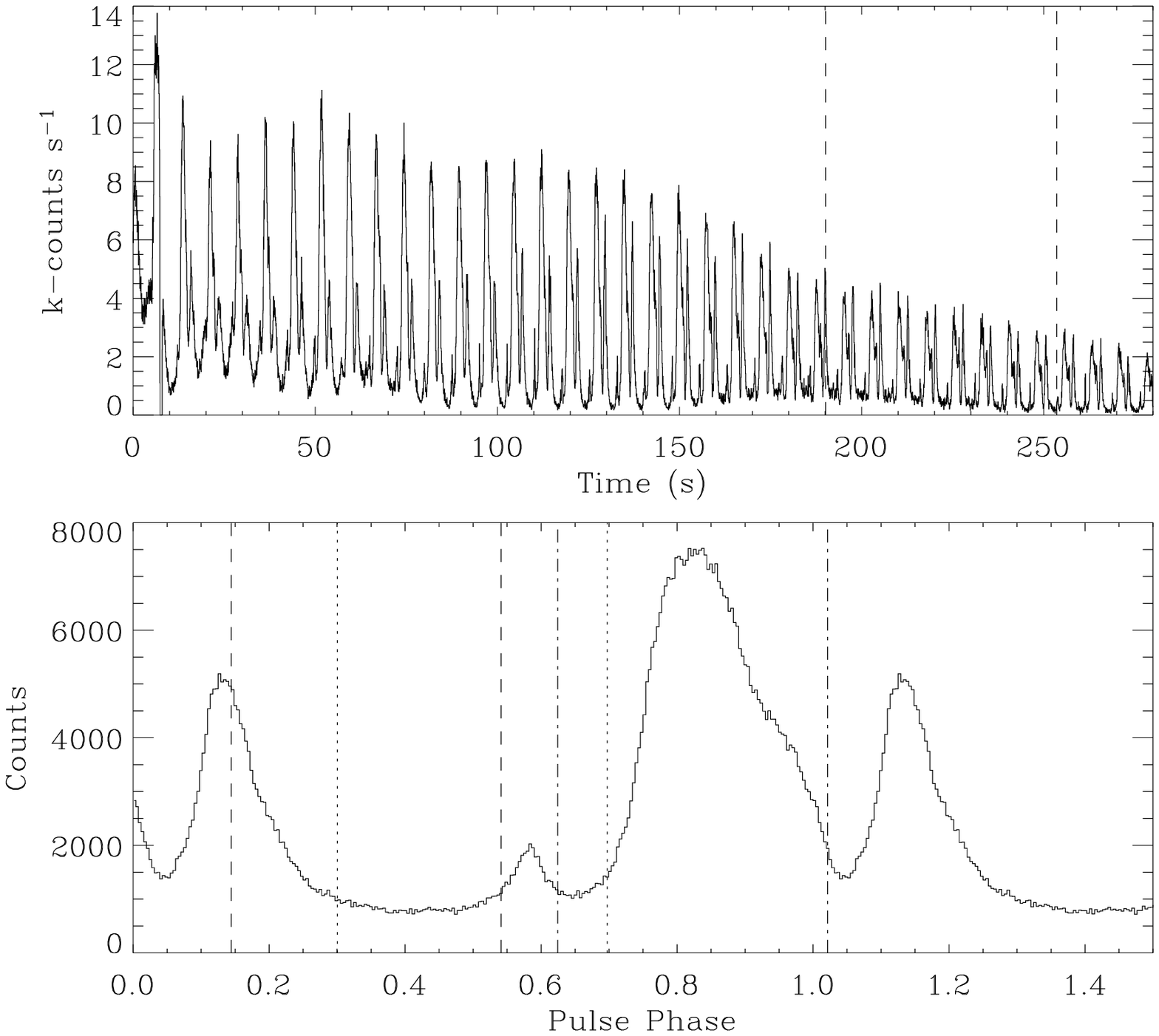}
\end{center}
\caption{X-ray intensity of the December, 2004 hyperflare as measured
by the PCA (top), and the average pulse profile (bottom). The curves
include all good events detected in PCA channels 10 - 200 (nominal
energy band from 4 - 90 keV). The main flare takes place approximately
4s prior to the zero on the time scale.  The time interval in which
the 625 Hz oscillation was detected is marked by the vertical dashed
lines (top).  The vertical lines in the bottom plot denote the phase
ranges used to compute power spectra shown in Figures 2, 4 and 5
(dashed), 6 (dash-dot), and 7 (dotted).  The text refers to Peaks 1, 2
and 3: Peak 1 is at phase $\approx$ 0.8, Peak 2 at $\approx$ 0.1 and
Peak 3 at $\approx$ 0.6.}
\label{fig1}
\end{figure}

\begin{figure}
\begin{center}
\includegraphics[width=6in, height=6in, clip]{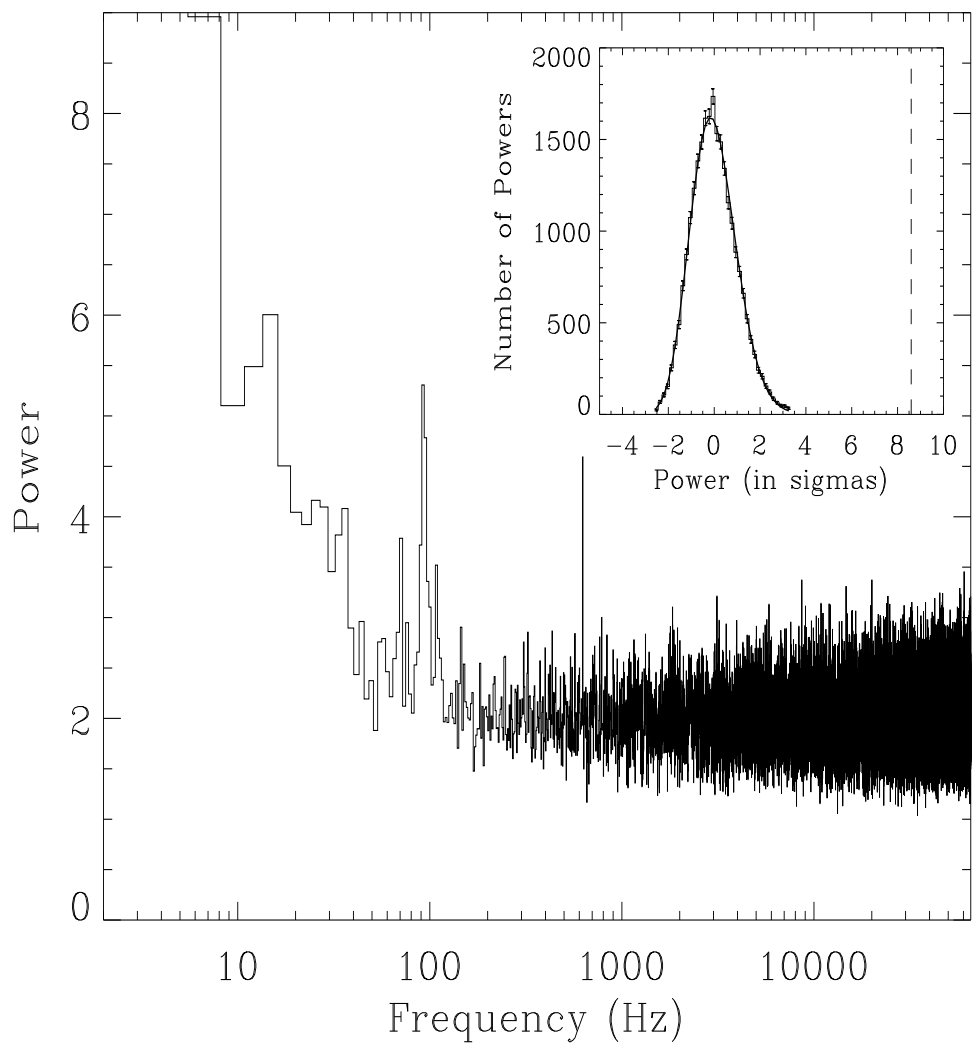}
\end{center}
\caption{Average power spectrum from a portion of the hyperflare from
SGR 1806-20 (main panel). We averaged nine 3 s power spectra from the
time interval marked by the vertical dashed lines in Figure 1
(top). The frequency resolution is 2.66 Hz. The inset panel shows the
distribution of noise powers computed from the frequency range 800 -
65,536 Hz as well as the best fitting $\chi^2$ distribution (solid).
The distribution is plotted in units of $\sigma$'s.  The vertical
dashed line marks the peak power of the 625 Hz feature.  See the text
for further discussion.}
\label{fig2}
\end{figure}

Figure \ref{fig2} shows an average power spectrum computed from 9
successive rotations starting approximately 190 s after the onset of
the hyperflare. Both the time interval during the flare (top), and the
phase region (bottom, dashed lines) used in computing this spectrum
are shown in Figure \ref{fig1}.  This power spectrum has a prominent
feature at 625 Hz, and the QPO at 92 Hz is also clearly detected. To
estimate the significance of the 625 Hz feature we extended the
frequency range to 65 kHz, and used all frequency bins above 800 Hz to
estimate the noise power distribution.  Figure \ref{fig2} shows the
power spectrum to 65 kHz (main panel), as well as the distribution of
noise powers (inset panel).  We fit the noise power histogram with a
$\chi^2$ function, and found a reasonable fit for 87 degrees of
freedom (dof), and a small reduction in the Poisson level of 0.01.
This function is also plotted in the inset panel of Figure \ref{fig2}
(solid curve). Using this noise distribution we find a single trial
significance of $7 \times 10^{-11}$ for the 625 Hz feature. The power
spectrum has 1,536 bins up to 4096 Hz (2.66 Hz resolution), which was
the top end of our search range.  We emphasize that we extended the
frequency range only {\it after} the search, simply to better
characterize the noise power distribution.  This gives a probability
of $1.1 \times 10^{-7}$ to find a peak this high in the power
spectrum. The spectrum shown in Figure \ref{fig2} was one of a
sequence from across the entire duration of the flare. We computed
these by overlapping the time intervals, so each spectrum in the
sequence is not fully independent.  However, even using the total
number of spectra computed (47), we have a significance $< 1 \times
10^{-5}$, so this is a robust detection.  We fit the QPO with a
Lorentzian profile and find a center frequency and width of $625.5 \pm
0.15$ Hz, and $1.8 \pm 0.4$ Hz, respectively.  The 625 Hz QPO has an
average amplitude during this interval of $8.5 \%$. We further note
that the strength of the 625 Hz feature is comparable to that of the
92 Hz signal, which provides additional confidence that it is not
simply a statistical fluctuation.

We searched the RHESSI data to see if there was any indication of a
simultaneous 625 Hz peak but did not detect anything significant at
this time.  This is however not surprising given the amplitude, and
the fact that for such high frequencies we can only use the RHESSI
front segments, which have a much lower countrate than RXTE (see the
discussion in \citet{wat06}).  In addition the RXTE signal is
strongest at energies less than 45 keV; at low energies the RHESSI
spectrum exhibits much higher background levels, which could swamp
such a weak signal.

\begin{figure}
\begin{center}
\includegraphics[width=6in, height=6in, clip]{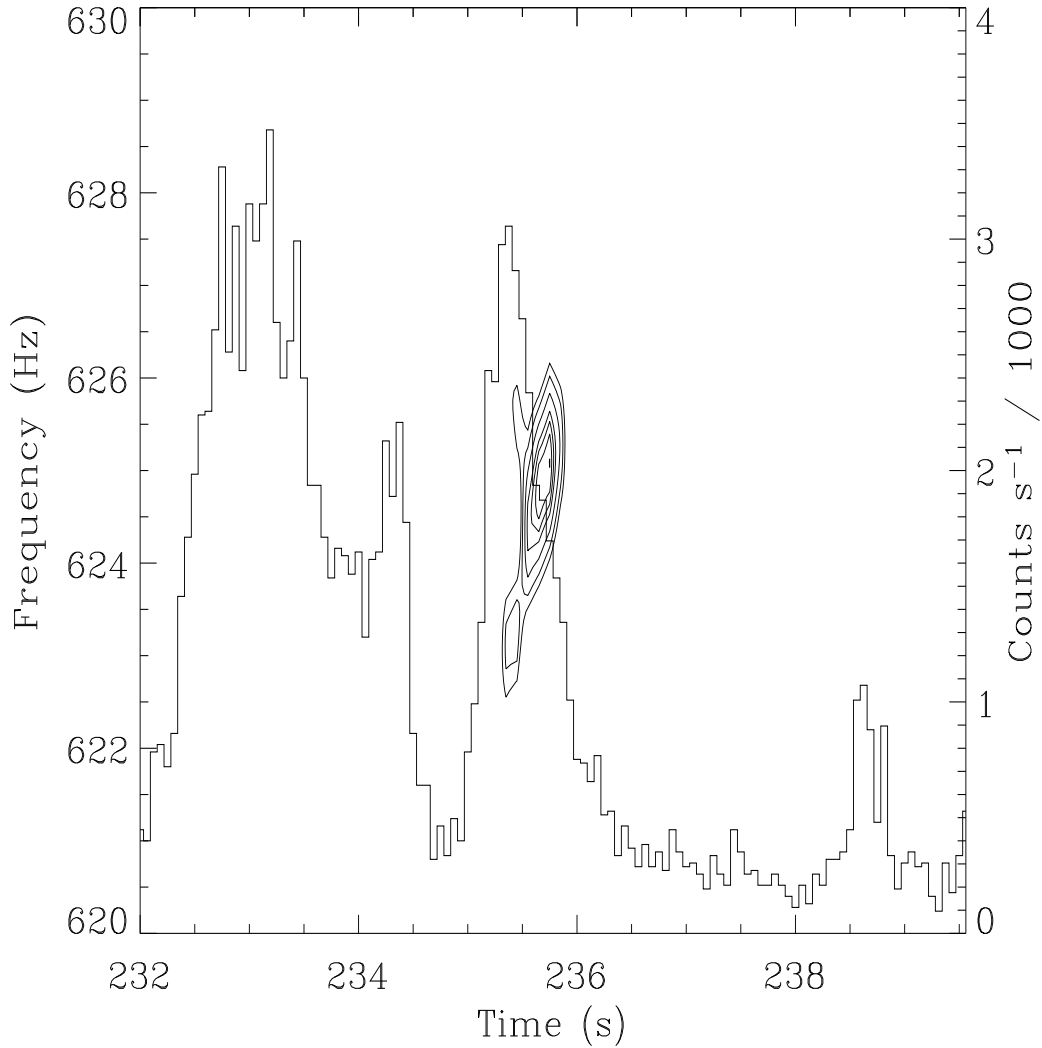}
\end{center}
\caption{Dynamic power spectrum of a portion of the data interval in
which the 625 Hz oscillation was detected. Contours of constant
Fourier power are plotted as a function of frequency and time, along
with the X-ray intensity as a function of time. We used 0.5 s
intervals to compute the power spectra, and overlapped the intervals,
beginning a new one every 0.1 s.  A strong contribution to the 625 Hz
signal detected in the average power spectrum of Figure 1 is evident
on the falling edge of peak 2.  Contours at Leahy normalized power
levels of 16, 18, 20, 24, 28 and 32 are shown. A single rotational
cycle is plotted.}
\label{fig3}
\end{figure}

We computed dynamic power spectra in the vicinity of 625 Hz to search
for time dependence of the signal.  The results suggest that the 625
Hz oscillation is most strongly associated with the falling edge of
peak 2 of the pulse profile, and that it's amplitude can vary strongly
with time. For example, Figure \ref{fig3} shows a portion of the
hyperflare that contributes strongly to the 625 Hz signal detected in
the average power spectrum described above.  Contours of constant
Fourier power in the vicinity of 625 Hz are plotted over the
hyperflare profile. We used 1/2 s intervals and started a new interval
every 0.1 s.  One pulsation cycle is plotted. The strong correlation
of power contours with the falling edge of peak 2 is rather similar to
the behavior seen in the 84 Hz oscillation from SGR 1900+14
(Strohmayer \& Watts 2005, see their Figure 3), and these similarities
suggest it may be connected with the X-ray modulation mechanism.

\begin{figure}
\begin{center}
\includegraphics[width=6in, height=6in,clip]{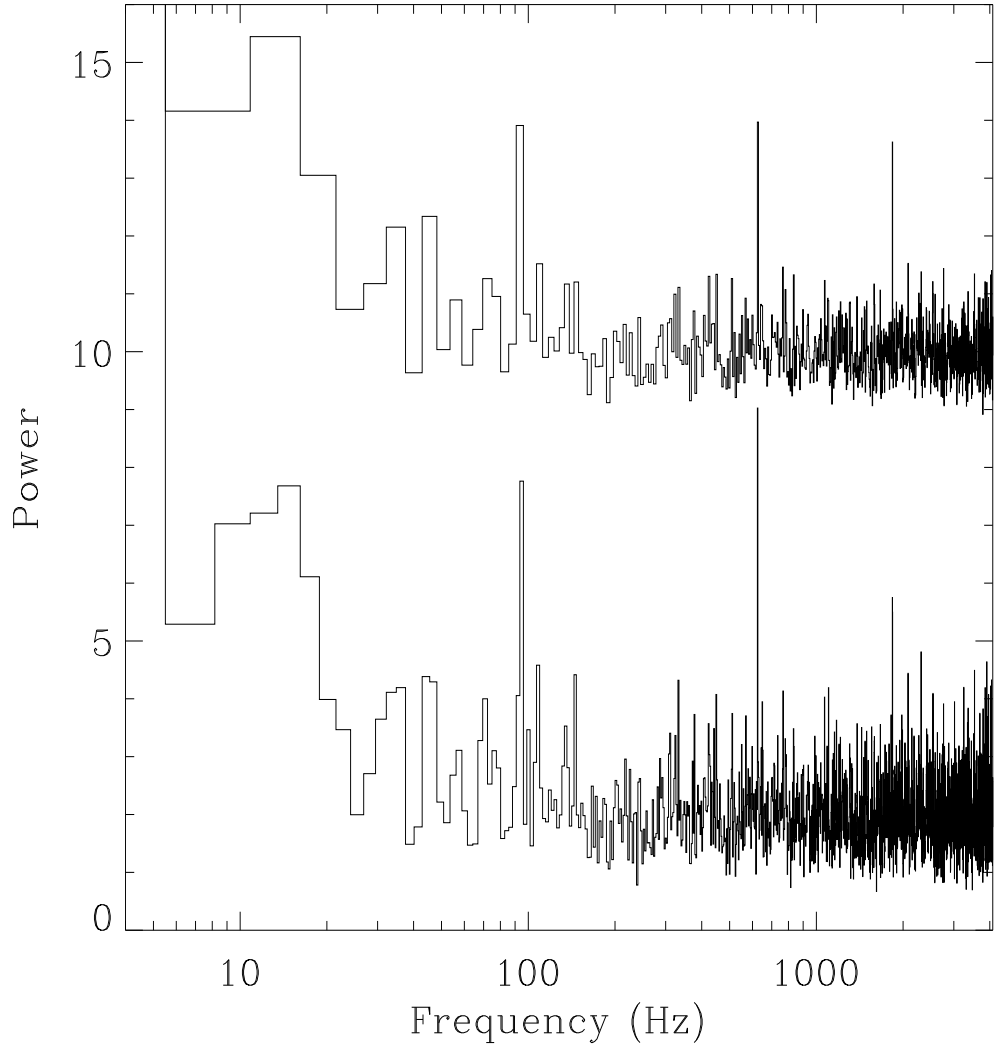}
\end{center}
\caption{Average power spectrum from two consecutive cycles beginning
with the cycle immediately prior to the pulse displayed in Figure 3
which shows a strong 625 Hz signal.  Two representations of the same
power spectrum are shown, only the frequency resolutions differ. The
frequency bins are 2.667 Hz wide in the bottom trace and twice that in
the top. Three frequencies are prominent in the top trace, 92 Hz, 625
Hz, and 1,840 Hz.  See the text for additional discussion.}
\label{fig4}
\end{figure}

We also computed several average power spectra in the immediate
temporal vicinity of the strong 625 Hz signal seen in the dynamic
power spectrum.  An average from two consecutive rotation cycles
beginning with the cycle prior to that shown in Figure \ref{fig3}
reveals evidence for an oscillation at an even higher frequency of
1,840 Hz. The average power spectrum from these two cycles is shown in
Figure \ref{fig4}. The two spectra are the same except for being
plotted at different resolutions.  We reiterate that the phase range
used here (and for Figure \ref{fig5} below) is the same as for Figure
\ref{fig2} (denoted by the vertical dashed lines in Figure
\ref{fig1}). The peak at 1,840 Hz is most prominent in the top trace,
at 5.33 Hz resolution (the bottom trace has 2.66 Hz resolution). Both
the 92 Hz and 625 Hz peaks also stand out prominently in this
spectrum.  We estimated the significance in the manner discussed
above, and found chance probabilities of $1.1 \times 10^{-6}$, and
$1.1 \times 10^{-5}$ for the 625 and 1,840 Hz peaks, respectively.
This is an estimate of the probability of finding a peak as high as
those observed in this power spectrum up to 4096 Hz. We computed eight
such spectra, so this will reduce the significances somewhat, but
again, the detections are further supported by the comparable
amplitude of these peaks to that of the well established 92 Hz
oscillation, which is also detected in the same power spectrum.  The
average amplitudes for the 625 and 1,840 Hz oscillations over these
two cycles are 18.3 and 18.0 \% (rms) respectively.

We also see weaker indications for a few additional high frequency
peaks during this time interval.  Figure \ref{fig5} compares two power
spectra, one (top) from the two cycle interval in which the 1,840 Hz
signal is seen, and the other (bottom) an average of two cycles
beginning six cycles earlier than the cycle shown in Figure
\ref{fig3}. The frequency resolution is 10.66 Hz in both plots.  The
top spectrum is actually the same as in Figure \ref{fig4}, but plotted
with lower resolution.  The 625 and 1,840 Hz peaks still stand out
prominently.  We estimated the standard deviation in the noise
distribution as before, and we placed two horizontal dotted lines
4$\sigma$ (single trial) above the mean in each case.  In the bottom
spectrum, there are also two peaks which reach above the $4\sigma$
line. These have frequencies of 720 and 2,384 Hz. We estimate
significances (single trial) of $4.4 \times 10^{-6}$ and $1.3 \times
10^{-5}$, respectively, for these peaks. A third peak, at 976 Hz, does
not reach the 4$\sigma$ line, but is a $3\sigma$ deviation, with an
estimated significance of $7.6 \times 10^{-4}$.  While we do not claim
that all these peaks represent strong detections, it does seem
unlikely by chance to have several such sizable deviations from the
mean in a single spectrum, and thus we think this could be an
indication of the presence of additional, weaker signals in the data.

We also searched the RXTE data for evidence of a 626 Hz QPO earlier in
the tail, when a QPO of that frequency was seen in the RHESSI data
\citep{wat06}.  We were not able to find a corresponding signal in the
PCA data. However, this is not entirely surprising.  The 626 Hz signal
was only observed in the RHESSI data for photons with recorded
energies in the range 100-200 keV.  Whilst RHESSI is sensitive at
these energies, the PCA does not record photons with energies above
$\approx$ 100 keV.  Comparing spectra from the two spacecraft is
difficult because of the scattering caused by the oblique incidence
angles.  However we can get some idea of the magnitude of the energy
shift between the two datasets by comparing the positions of the
spectral maxima.  For the PCA the maximum is at 20 keV, for RHESSI it
is at 40 keV.  If incident photons are downshifted in energy by only
20 keV for the RXTE PCA compared to RHESSI, then we would not expect
to see the RHESSI 626 Hz signal, since the PCA does not capture
sufficient high energy photons.  Unfortunately the RXTE High Energy
X-ray Timing Experiment (HEXTE), which covers a higher energy range
than the PCA, recorded the main flare but not the decaying tail
\citep{smi05}.
 
\subsection{150 Hz Oscillation}

We also computed average power spectra centered on the rotational
phase of peak 1.  Figure \ref{fig1} shows the range of phases used
(dashed-dot lines). Average power spectra centered on this phase over
the entire hyperflare show a significant QPO centered near 150 Hz.
Figure \ref{fig6} shows the power spectrum in which we detected this
QPO.  We averaged 45, 3 second power spectra, and the frequency
resolution is 8 Hz.  The peak near 150 Hz is evident.  We modelled the
power continuum in the range above 30 Hz with a constant plus a power
law, and then rescaled the spectrum by dividing by the best fitting
continuum model.  We then estimated the noise power distribution in
the same way as described earlier, by fitting a $\chi^2$ distribution
to a histogram of noise powers in the rescaled spectrum. Based on this
we estimate a significance for the highest bin in the QPO profile as
$2.8 \times 10^{-5}$, which includes the number of frequency bins
searched (up to 4096 Hz).  This is a very conservative estimate of the
QPO significance, because it only uses the highest bin, and the QPO
peak is resolved at this frequency resolution.  We fit the QPO profile
with a lorentzian and find a centroid frequency of $150.3 \pm 1.6$ Hz,
a width of $17 \pm 5$ Hz, and an amplitude (rms) of $6.8 \pm 1.3 \%$.
If we construct averages from the first and second halves of the
hyperflare, we still detect the 150 QPO, so this oscillation seems to
persist through most of the hyperflare. This feature is not detected
in the RHESSI data, but this is not unexpected given the low amplitude
of the QPO and the lower countrates of the RHESSI front segments.

\subsection{Lower frequency QPOs}

\citet{isr05} reported the marginal detection of 18 Hz and 30 Hz QPOs
over a period late on in the burst, without exploring for any phase
dependence.  \citet{wat06} then found evidence in the RHESSI dataset
for broad QPOs at 18 and 26 Hz, with a weaker feature at 30 Hz.  These
QPOs showed a strong rotational phase dependence, appearing at the
same phase as the 92.5 Hz QPO.  For these lower frequency QPOs we were
able to include photons detected by the RHESSI rear segments, so that
the RHESSI countrate exceeded that recorded by RXTE.

The broad QPOs at 18 Hz and 26 Hz were seen in the RHESSI dataset
between 60 and 230s after the main flare.  We were interested to see
whether these QPOs were detectable in the RXTE data.  Folding up
exactly the same time period and rotational phase used in the RHESSI
analysis we were unable to make a statistically significant detection,
not unexpected given the discrepancy in countrates and the low
fractional amplitude of the QPOs.  We therefore focused on two shorter
periods in the RHESSI data when the 18 Hz and 26 Hz oscillations were
particularly strong, and searched for them at corresponding times in
the RXTE data.  In both of these cases corresponding peaks are
observed in the RXTE data, confirming the detection of QPOs at these
frequencies (Figure \ref{fig7}).  Interestingly the 26 Hz QPO was
stronger in this period in the RXTE dataset than it is in RHESSI,
suggesting that other factors may offset the higher countrate (see
below).

What about the 30 Hz QPO?  If we consider the same time interval
indicated in Figure \ref{fig1}, but shift the phase range over which
we compute power spectra to later phases, that is, exclude the falling
edge of peak 2, the 625 Hz signal drops to undetectable levels. Figure
\ref{fig8} shows the average power spectrum of this ``interpulse''
region during this time interval, and confirms the initial claims of
Israel et al. (2005) that signals at $\approx 30$ and 92 Hz are
present towards the end of the hyperflare.  These QPOs are extremely
significant, for example, we conservatively estimate the significance
of the 30 Hz QPO at $1.4 \times 10^{-11}$, and the 90 Hz feature is
even more significant. These QPOs can be fitted with lorentzian
profiles. Based on such fits we find centroid frequencies of $28.98
\pm 0.4$, and $92.9 \pm 0.2$ Hz; quality factors ($\nu_0 / \Delta\nu$)
of $7 \pm 0.8$, and $39 \pm 5$; and average amplitudes of $20.5 \pm 3$
$\%$, and $19.2 \pm 2$ $\%$, respectively.

The 30 Hz feature in the RXTE spectrum is far more significant than
the weaker feature at 30 Hz found in the RHESSI spectrum.  This is at
first glance surprising, since RHESSI has a higher countrate once the
rear segments are included.  There are however several factors that
probably contribute to this discrepancy.  Firstly, we are in the
interpulse region, where RHESSI's high background levels are more
important.  Secondly, one cycle of this time period was excised from
the RHESSI data because of an artificial spike in countrate as a
protective attenuator was removed. In addition, as discussed above, we
find occasions where the 26 Hz QPO appears far more prominently in the
RXTE data than in the RHESSI data, suggesting that other factors such
as scattering off the spacecraft and the Earth may reduce RHESSI's
effectiveness and offset the higher countrate.

\subsection{Amplitude variation and frequency drift}
\label{drift}

In the periods when they are detected, the QPOs are far from static.
Their amplitudes wax and wane, and there is evidence for frequency
shifts and possible multiplet splitting.  In Figure \ref{fig9} we show
a sequence of dynamical power spectra for the Peak 2/Interpulse
region, for the time period when the 92.5 Hz QPO is active.  The
presence of the strong QPO at this frequency from 180-230s after the
main flare is clear (Panels 17-24 of Figure \ref{fig9}). If we look at
the cycles around this period however, there are hints of interesting
behavior.  Back at only 80s after the flare, weak detections are made,
with a frequency that starts at around 78 Hz and rises up to a strong
92.5 Hz QPO that persists for a couple of cycles from 120-130 s.  The
QPOs then weaken, and the frequency seems to fall again by a few Hz,
until the strong 92.5 Hz signal reappears at 180s.  At later times, as
already noted by \citet{isr05}, the frequency rises to 95 Hz and
possibly higher, as the amplitude falls and the signal becomes
undetectable.

The variation of amplitude over the rotational cycle, and the
consequent interruption of the signals, means that we cannot
rigorously connect cycles, as one would wish to in order to verify
that the frequency is evolving.  However, our results are suggestive
of this eventuality.  There is also some suggestion of frequency
movement in the lower and higher frequency QPOs (see for example
Figure \ref{fig3}), although this is even harder to track from cycle
to cycle because the QPOs are weaker and the amplitudes more variable.
As such it is not possible to verify whether the other QPO frequencies
move in concert with the 92.5 Hz QPO.

\section{Discussion}

Using phase averaging techniques we have found evidence for
additional, higher frequency QPOs during the December, 2004 hyperflare
from SGR 1806-20. The properties of the QPOs, reported in this and
previous papers, are summarized in Table \ref{qposum} and Figure
\ref{fig10}.  The additional frequencies and the time and phase
dependence of the detections suggest an extremely rich and complex
phenomenology associated with the magnetar oscillations.

The detection of a strong 625 Hz QPO in the RXTE data is particularly
interesting. The frequency is remarkably similar to that of a QPO
found by us in the RHESSI data, but there are some intriguing
differences in the properties of the two QPOs. In addition to having a
slightly lower centroid frequency (there is frequency overlap of the
QPO profiles however), the RXTE QPO has substantially lower fractional
amplitude and coherence value $Q$.  The RXTE QPO appears later in the
tail of the flare, and the apparent photon energies of the two QPOs
also differ.

The RHESSI detection used events from the front segments, in the 100 -
200 keV energy band (see \citet{wat06} for a discussion of RHESSI data
in this context).  The phase interval was centered on the rising edge
of peak 1 (\citet{wat06}, Figure 4), and the average was obtained from
about 50 to 200 s after the start of the hyperflare.  An average power
spectrum from RXTE for the same time and phase ranges shows no
corresponding signal in the PCA.  However, it is difficult to know
exactly how to compare the energies of events detected in the two
instruments. For RHESSI, many of the events observed in the front
segments were direct (ie. meaning that they were not scattered before
detection), wheareas this is likely not the case with RXTE, as
virtually no photons would have made it directly into the PCA xenon
volumes without having scattered off of some part of the
spacecraft. It is quite possible that no photons with incident
energies in the 100 - 200 keV range were detected by the PCA, which
would account for the non-detection.  The later, lower amplitude RXTE
QPO is not detected in the RHESSI data, most probably due to the lower
countrates.

The lack of simultaneous detections is therefore disappointing, but
not surprising.  Perhaps more important is the fact that the two
instruments detected signals independently at consistent frequencies.
The QPO profile of the 625 Hz feature detected with RXTE overlaps the
1 Hz bin (centered on 626.5 Hz) in which the signal was detected with
RHESSI. Although it is difficult to determine an {\it a priori}
probability, it seems extremely unlikely that both instruments would
detect signals at a consistent frequency unless that frequency were
intrinsic to the source.  We therefore think there is compelling
evidence to associate this frequency with SGR 1806-20.

The finding of additional characteristic frequencies in the SGR
1806-20 hyperflare further strengthens the connections between the
oscillations seen in both the SGR 1806-20 and SGR 1900+14 events.  In
each case, oscillations near $\approx 30$, $\sim 90$ and $\approx 150
$ Hz have been seen. This strongly suppports the notion that the same
physical processes are involved in these objects.

Several mechanisms have been suggested as the cause of the QPOs.  One
suggestion is that the QPOs are caused by an interaction with a
remnant or ejected debris disk, the mechanism being similar to that
which gives rise to the kHz QPOs in accreting neutron stars. Although
there is now evidence that some magnetars have disks \citep{wan06},
there is no evidence as yet for a disk around the SGRs, and it is also
very difficult to understand the rotational phase dependence if a disk
is responsible.

An alternative is that we are seeing oscillations of the plasma-filled
magnetosphere and the trapped fireball. Simple estimates of the
Alfv\'en speed in the magnetosphere suggest that magnetospheric
frequencies would be too high to explain the observations, but
evolution of the magnetosphere and fireball over the course of the
tail might be a natural explanation for the frequency drift suggested
in Section \ref{drift}. The theoretical details of such modes remain
to be worked out, however, and we will not discuss the magnetospheric
model further in this paper.  We will instead focus on the third and
perhaps most promising mechanism, neutron star vibrations.

\subsection{Neutron star oscillations}

The detection of similar frequencies in SGR 1806-20 and SGR 1900+14
makes sense in the context of global oscillation modes; if the neutron
stars have similar masses and magnetic fields, then they should ring
with the same set of characteristic frequencies.  The detection of
similar frequencies at different rotational phases (the $\approx$ 625
Hz QPOs) also argues in favor of an underlying global oscillation,
with magnetospheric beaming/obscuration effects causing the phase
dependence. Frequency drift can also be incorporated within this
model, as discussed later.

Attention has so far focused on the toroidal shear modes of the
neutron star crust, thought to be both easy to excite and strongly
coupled to the external magnetosphere (providing a means to modulate
the X-ray lightcurve).  The most detailed computations of mode
frequencies to date have included both the depth-dependence of the
shear modulus within the crust, and the effective boost to the shear
modulus due to the magnetic pressure \citep{dun98, pir05}.
Potentially important effects that have not been addressed in detail
include coupling between crust and core due to the magnetic field, the
equation of state in the boundary layer between crust and core, and
non-uniformity of the magnetic field.  In what follows we will refer
to the models that assume no coupling between crust and core as ``pure
crust'' models; and those that do include coupling due to the field as
``coupled crust-core'' models.

The question of whether or not ``pure crust'' models are adequate is
an important one.  \citet{lev06} used a toy model to show that if
there is a strong perpendicular magnetic field threading the boundary,
then large horizontal displacements at the base of the crust must
excite vibrations in the core.  The effects of coupling may however be
mitigated in several ways.  For example, coupling to the core will
depend sensitively on the amplitude at the crust core boundary, and
previous calculations show that this amplitude can be substantially
smaller than that at the top of the crust (see McDermott, van Horn \&
Hansen 1988, for example).  Thus, approximating the amplitude as a
single constant for a given mode, as assumed by Levin, may not be very
accurate.  The physics at the boundary layer at the base of the crust
will also be crucial (see for example \citet{kin03}).  The elastic
properties at the base of the crust \citep{pet98} have not yet been
considered, and could substantially modify the eigenfunctions in this
region.  In addition, coupling will likely depend rather sensitively
on both the magnetic field geometry and the particular displacement
pattern of individual modes.  The modes that persist, and that we
observe, may be those for which the coupling is minimal.  The presence
of a strong toroidal field in the core of the star could also reduce
coupling by making core more rigid and less prone to excitation.

We will return to the issue of coupling later in this Section.  For
the moment, however, we will start by assessing the observations in
the light of the ``pure crust'' models.

\subsection{Interpretation as ``pure crust'' modes}

The 30 Hz, 92 Hz and 150 Hz QPO can be interpreted within the global
oscillation model as $n=0$, $l$=2, 7 and 12 torsional modes of the
neutron star crust.  \citet{wat06} suggested that the 626 Hz
oscillation seen in RHESSI data could be associated with a toroidal
mode having a single node in the radial displacement eigenfunction
($_l t_1$ modes).  Recent estimates from Piro (2005, see his Figure 3)
suggest that this frequency is consistent with theoretical
expectations.  Our confirmation of the same frequency in the RXTE
data, provides a firm indication that we are seeing not only modes
with a sequence of different $l$-values, but also modes with different
$n$-values.

It is possible that the 625 Hz QPO seen in the RXTE dataset represents
an evolution of the QPO seen earlier in the flare in the RHESSI data.
However, the mode model also offers a possible explanation for the
detection of two $\approx$ 625 Hz QPOs with differences in
properties. Calculations (see \citet{HC80, pir05}), indicate that the
$n=1$ mode frequencies are not very sensitive to $l$. As such we could
be seeing two $n=1$ modes of different $l$, excited at different times
by either aftershocks or mode coupling.  The implications of detecting
an $n=1$ mode are discussed in more detail in Section \ref{thick}.

The 18 Hz and 26 Hz QPOs are at first glance too low in frequency to
fit simple torsional mode models, however, mode calculations with
realistic magnetic fields will have $m$-dependent splitting, perhaps
producing an observable multiplet structure. This might accommodate
the observed spread from about 26 - 30 Hz in a putative $_2t_0$ mode,
but detailed calculations will be required to explore this
quantitatively.  The 18 Hz QPO, however, remains difficult to
understand.

The stronger high frequency QPOs at 625 and 1840 Hz can be nicely
accommodated within the torsional mode hypothesis as low $l$, $n=1$
and $n=3$ modes, respectively (see, for example, Figure 3 in Piro
2005).  The less significant features at 720, 976, and 2,384 Hz, may
also fit within this scenario.  The 720 Hz might represent a low $n$
mode with significantly different $l$, as pointed out by Hansen \&
Cioffi (1980), the $l$ dependence of such modes is not strong, but can
produce a scatter in frequency on the order of 15\%.  The other
frequencies might correspond to modes with $n=2$, and $n > 3$,
respectively.  In addition, at these high frequencies, other modes,
such as the crustal spheroidal modes (see McDermott et al. 1988), may
become relevant. Again, more detailed mode calculations, including the
effects of magnetic fields, will be needed to make more quantitative
mode identifications.

\subsubsection{Implications for Neutron Star Crusts}
\label{thick}

If the 625 Hz QPOs are indeed $n=1$ torsional modes, the implications
for the study of neutron star structure are profound.  The ratio of
frequencies of the $n=1$ and $n=0$ torsional modes enables one to
deduce the depth of the neutron star crust (see Hansen \& Cioffi 1980;
Piro 2005).  The basic physics behind this estimate is that the $n=0$
mode frequencies are set by the horizontal wavelength, which scales
with the stellar radius, $R$, whereas the $n>0$ mode frequencies are
dominated by the vertical wavelength, which senses the depth of the
crust, $\Delta R$.  In the limit of a thin crust and a constant shear
wave speed, one finds the simple expression (see Hansen \& Cioffi
1980; McDermott et al. 1988; Piro 2005),
\begin{equation}
\frac{\nu_{l, n=0}}{\nu_{n>0}} = \frac{\left ( l(l+1) \right
)^{1/2}}{3 n} \frac{\Delta R}{R} \; .
\end{equation}
If we identify 30 Hz with the $l=2, \; n=0$ mode frequency, and 625 Hz
with the $n=1$ mode (and $l \approx n$), the above expression predicts
$\Delta R/R = 0.06$.  In most modern neutron star models, however, the
crust is not ``thin'' in the sense of $\Delta R /R << 1$, and the
shear speed has some variation with depth, so this simple estimate
should probably be considered a lower limit.  A more accurate estimate
can be obtained by comparisons with more detailed mode calculations.
For example, using the calculations of McDermott et al. (1988) we
evaluated the expression, $(\nu_{l,1} \Delta R)/ (\nu_{2,0} R)$, that
according to the approximate theory should be the proportionality
constant between the mode frequency ratio and the fractional depth of
the crust.  For the four models tabulated by McDermott et al. (1988)
we find that this expression ranges from 2.15 to 2.65.  Adopting this
range as representative, and using the scale factors to convert our
observed mode frequency ratios into depths, we obtain the range, $0.1
< \Delta R / R < 0.127$. We note that the models from McDermott et al.
use softer equation of state models than are currently favored, so it
is important to repeat this exercise with more up to date neutron star
models, and also incorporate magnetic fields.  In addition the effects
of possible crust-core coupling must be considered (see next
subsection).

Measuring the thickness of a neutron star crust also conveys
information on the equation of state (EOS). In other words, stellar
models of a given mass computed with different EOSs will in general
have crusts of different depths.  For example, Lattimer (2006) has
recently outlined how constraints on the fractional depth and stellar
mass could be used to constrain the nuclear symmetry energy and the
nuclear force model.  The crust depth dependence on the EOS also has
implications for pulsar glitch models (see, for exmaple, Crawford \&
Demianski 2003; Link, Espstein \& Lattimer 1999).

\subsection{Coupled crust-core modes}

If the coupling between crust and core is strong, by virtue of
boundary layer physics or magnetic field threading, one needs to
consider the global magneto-elastic modes of the neutron star.  As
first suggested by \citet{isr05}, global modes can accommodate the
lower frequency 18 and 26 Hz QPOs very easily.  A recent paper by
\citet{gla06} developed a simple slab model of global magneto-elastic
oscillations that showed two interesting features.  Firstly, it
confirmed the presence of modes at lower frequencies than the ``pure
crust'' toroidal modes.  Secondly, the model exhibited modes for which
the amplitudes in the crust were strong; in these cases the frequency
was very close to the well-established ``pure crust'' frequencies. In
other words, even with coupling included, it is possible to obtain
very similar frequencies to those that we know match the
data. Although the model was very simple (slab geometry, for example,
is not adequate to describe behavior deep in the core), this gives us
some indication that global modes may have similar frequencies.  Low
core amplitudes would also reduce damping, another issue of concern
for \citet{lev06}\footnote{The toy model developed by Levin possesses
a continous spectrum of apparently singular modes in the core fluid
that he claims will drain energy from the system, leading to rapid
damping of global oscillations.  Firstly, the continuous spectrum in
his problem is a consequence of the unbounded geometry and simplified
assumptions of the toy model; whether such a continuum would exist in
a realistic stellar model is not yet clear.  Secondly, an apparently
singular continuous spectrum of eigenfunctions can form a non-singular
collective perturbation when one considers the initial value problem
(the response to an initial perturbation such as a crust fracture).
The temporal behavior and response to damping of such a perturbation
is far from straightforward \citep{wat03, wat04}.  As such we think
Levin's conclusions regarding the lifetimes of global oscillations are
premature.}. The bounds on crust thickness estimated in section 3.2.1
above were based on ``pure crust'' models.  In principle, however, a
similar calculation could be done for coupled crust-core modes, using
the ratio of two frequencies to scale out gravitational redshift
effects.

%\subsection{Implications for Neutron Star Crusts}
%\label{thick}

\subsection{Temporal variation}

Figure \ref{fig10} shows the periods during the tail of the flare when
the different QPOs are detected in either the RXTE or RHESSI datasets.
Some of the oscillations are seen throughout the flare, others are
detected only after the rise in emission half way through the tail,
and the highest frequency modes are very short-lived.  This complex
temporal variation requires explanation, and to do this we must
understand the excitation and damping mechanisms.

The data suggest that some QPOs are excited by the main flare, whereas
others are excited later in the tail, perhaps by aftershock fracturing
and reconnection, or energy release from the core.  The association of
several of the QPOs with the rise in emission late in the tail lends
support to the idea of continued excitation.  However, there remains
the possibility that all of the QPOs are excited by the main flare,
with their visibility depending on beaming effects that vary during
the tail. Beaming will be influenced by both the magnetic field and by
the plasma surrounding the magnetar.  Selection by beaming might also
account for the fact that only some of the myriad oscillations
predicted by theory are observed - although mode excitation will also
depend sensitively on the speed and extent of the fracture.  Some
modes may also damp far too quickly to be detected, depending on local
conditions in the crust and magnetic field geometry.  We believe that
these factors are more than capable of accounting for the fact that
only a few frequencies reach detectable amplitudes, an issue of
concern to Levin (2006).

Once a mode is excited, there is still significant variation in
amplitude and perhaps in frequency (as discussed in Section
\ref{drift}).  One possibility to explain amplitude variation is that
the mechanism which modulates the X-ray flux is strongly ``tunable.''
That is, at times when certain frequencies are detected their
instantaneous amplitudes can be quite high, almost certainly higher
than the relative surface motions.  So, it may be that the modulation
mechanism can amplify relatively small surface motions.  Resonant
interaction between the crust and the core could also lead to
amplitude variations over the course of the tail.  Frequency drifts
could arise in several ways.  Evolution of the magnetic field in the
aftermath of the flare, relaxation of the deep crustal structure, and
the evolution of the surrounding plasma, are all candidates for
causing frequency shifts.  Alternatively we could be seeing different
members of a magnetically-split multiplet excited, perhaps, by mode
coupling.
 
\section{Conclusions}

Our study of the RXTE data from the SGR 1806-20 hyperflare indicates
that a complex pulsation phenomenology is associated with magnetar
hyperflares.  The discovery of new kHz-range frequencies consistent
with theoretical predictions for $n>0$ torsional modes provides strong
evidence that we may be seeing vibration modes of the neutron star
crust excited by these catastrophic events. If this is true, then it
opens up the exciting prospect of probing the interiors of neutron
stars in a manner analogous to helioseismology.  Additional excitement
is warranted when we consider that all the current datasets used to
explore these oscillations have been purely serendipitous.  That is,
they have not been optimized in any way for studying these signals.
This suggests that a wealth of additional information would likely be
found from instruments better optimized to capture with high time
resolution the flood of X-rays produced by these events.  Based on the
findings presented here we are also strongly convinced that more
theoretical work is definitely needed to make more accurate mode
identifications, to better understand the excitation and damping
mechanisms of modes and how they can couple to the X-ray emission, and
to make more precise inferences on neutron star structure.

\acknowledgements

ALW acknowledges support from the European Union FP5 Research Training
Network ``Gamma-Ray Bursts: An Enigma and a Tool''.  TES thanks NASA
for its support of high energy astrophysics research.  We are also
grateful to the anonymous referee for a detailed report, including
many useful thoughts on QPO interpretation.

\clearpage

\pagebreak

\pagebreak

\pagebreak

\pagebreak

\begin{figure}
\begin{center}
\includegraphics[width=6in, height=6in,clip]{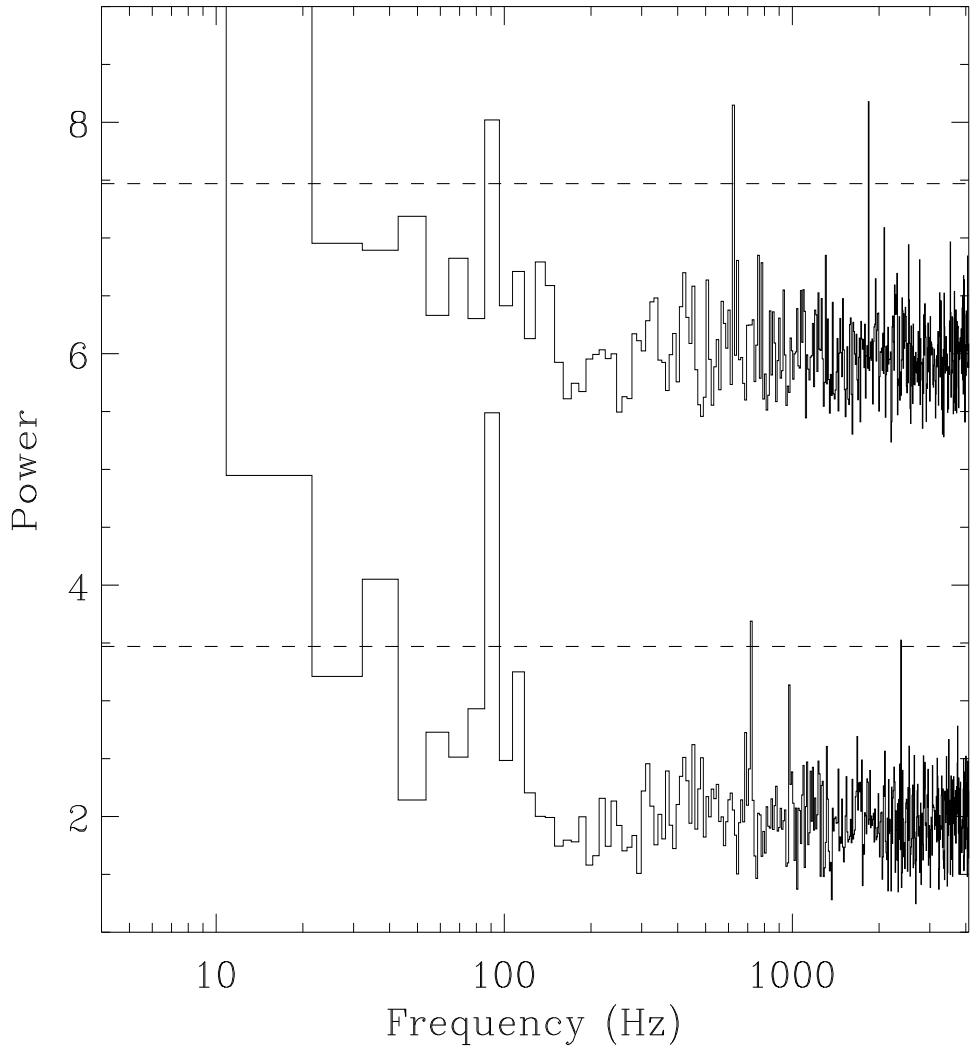}
\end{center}
\caption{Average power spectra from the time interval in which the 625
Hz oscillation was first detected. Shown are average power spectra
computed from two consecutive cycles. The top panel includes the
pulsation cycle shown in Figure 3, and the preceeding cycle (same as
in Figure 4, only plotted with lower frequency resolution).  The
bottom panel shows the spectrum from an earlier, 2-cycle time
interval, beginning six cycles before that shown in Figure 3.  In each
case the frequency resolution is 10.66 Hz. In each spectrum a
horizontal dotted line marks a $4\sigma$ (one trial) deviation. The
625 and 1,840 Hz features are detected in the top spectrum, while
several different frequencies are suggested in the bottom
spectrum. See the text for more details.}
\label{fig5}
\end{figure}

\pagebreak

\begin{figure}
\begin{center}
\includegraphics[width=6in, height=6in, clip]{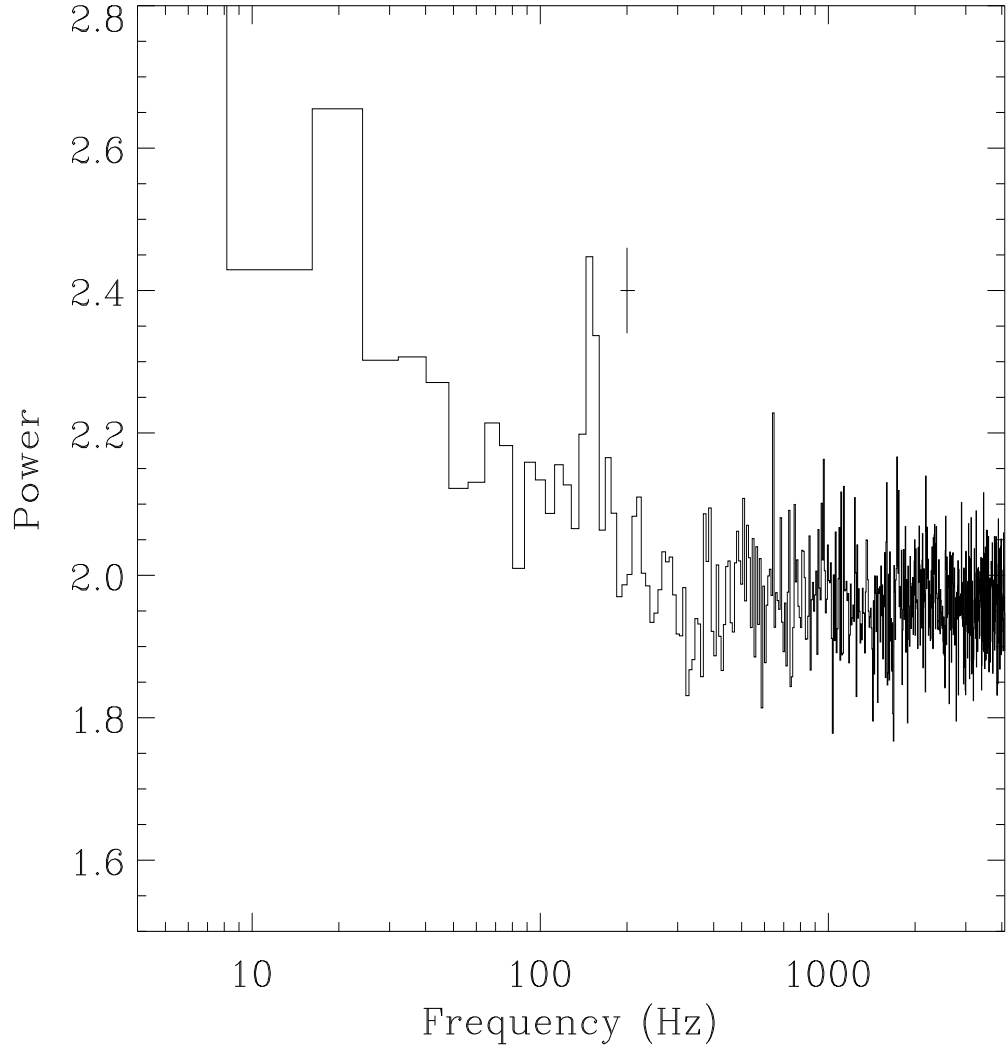}
\end{center}
\caption{Average power spectrum of the phase region centered on peak
1, showing the 150 Hz QPO. We averaged 45 power spectra, each computed
from 3 s of data over essentially the entire flare.  The phase
interval used is shown by the vertical dot-dashed lines in Figure 1. A
characteristic error bar is shown next to the QPO peak.}
\label{fig6}
\end{figure}

\pagebreak

\begin{figure}
\begin{center}
\includegraphics[width=6in, clip]{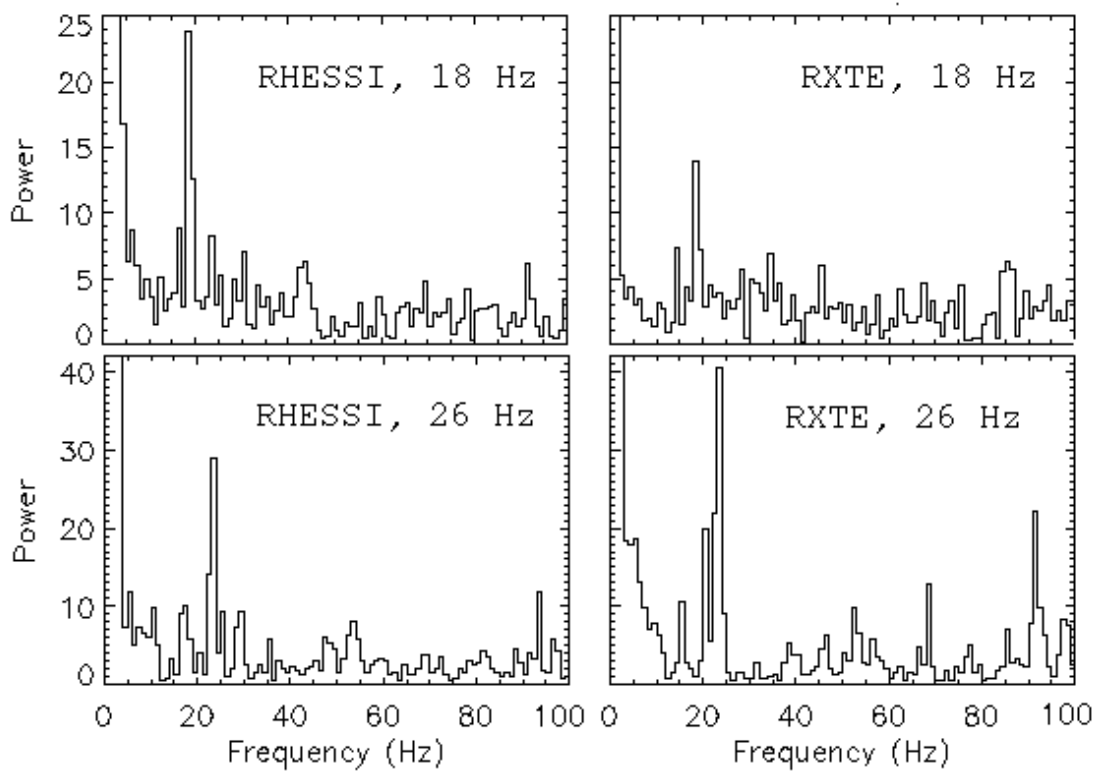}
\end{center}
\caption{Average power spectra computed for RXTE and RHESSI data, for
one cycle in which either the 18 Hz or the 26 Hz QPO is particularly
strong in the RHESSI dataset.  Each power spectrum has 1 Hz frequency
resolution, and is computed using 0.3 rotational cycles.  The interval
shown for the $\approx$ 18 Hz QPO starts 89s after the main flare;
that for the $\approx$ 26 Hz QPO Starts 127s after the main flare.
Simultaneous peaks are seen in the RXTE data despite the apparent
countrate disadvantages of RXTE.}
\label{fig7}
\end{figure}

\pagebreak

\begin{figure}
\begin{center}
\includegraphics[width=6in, height=6in,clip]{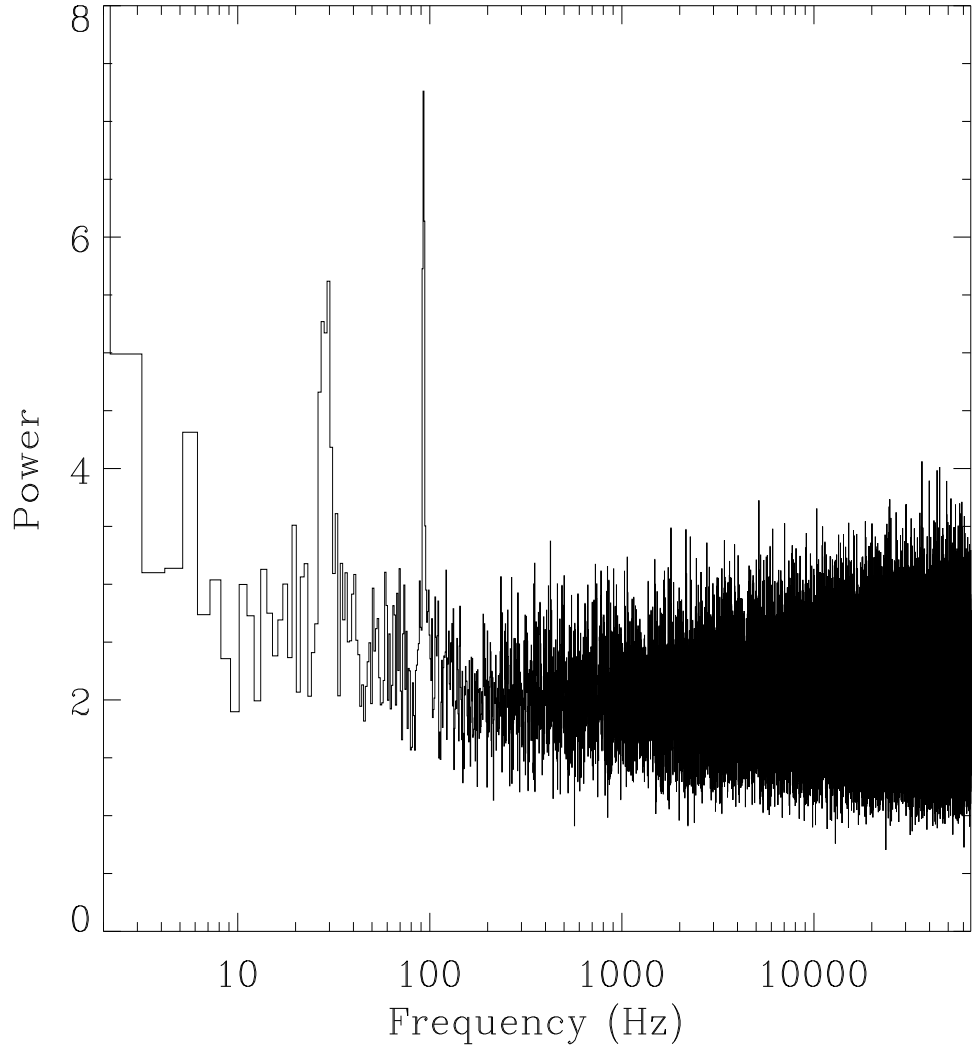}
\end{center}
\caption{Average power spectrum from the ``interpulse'' region during
the same time interval in which the 625 Hz oscillation was detected.
The phase interval used for the average spectrum is marked with
vertical dotted lines in Figure 1. Strong QPOs at 29 and 92 Hz are
clearly detected.  See the text for additional discussion.}
\label{fig8}
\end{figure}

\pagebreak

\begin{figure}
\begin{center}
\includegraphics[width=6.5in, clip]{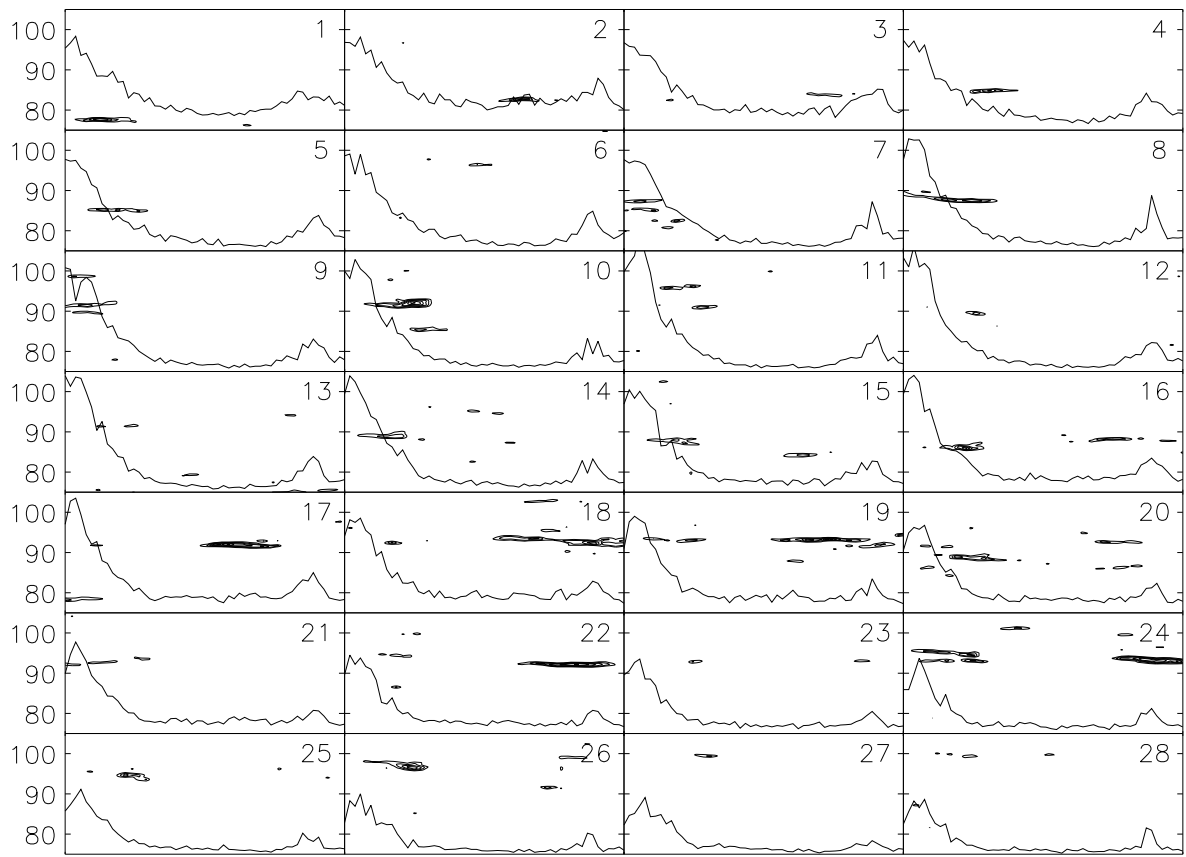}
\end{center}
\caption{The evolution of frequency in the 75-105 Hz band, for the
Peak 2/interpulse region. Each numbered panel shows a 4s interval for
consecutive cycles, the first panel starting $\approx$ 60 s after the
main flare. The y-axis shows frequency in Hz.  Dynamical power
spectra, shown as contours, are computed from 1s segments, overlapping
by 0.1s.  The minimum contour level shown is a power of 15, with
increments of 5. Lightcurves are also shown, plotted to the same scale
on each panel.  The plots show amplitude variation, frequency
evolution, and possible multiplet activity.  The start time of each
panel in seconds (relative to the initial flare at time zero) is: (1)
59.8, (2) 67.4, (3) 75.0, (4) 82.5, (5) 90.1, (6) 97.7, (7) 105.3, (8)
112.8, (9) 120.4, (10) 128.0, (11) 135.6, (12) 143.1, (13) 150.7, (14)
158.3, (15) 165.9, (16) 173.4, (17) 181.0, (18) 188.6, (19) 196.2,
(20) 203.7, (21) 211.3, (22) 218.9, (23) 226.5, (24) 234.0, (25)
241.6, (26) 249.2, (27) 256.8, (28) 264.3}
\label{fig9}
\end{figure}

\pagebreak

\begin{figure}
\begin{center}
\includegraphics[width=5.5in, angle=270, clip]{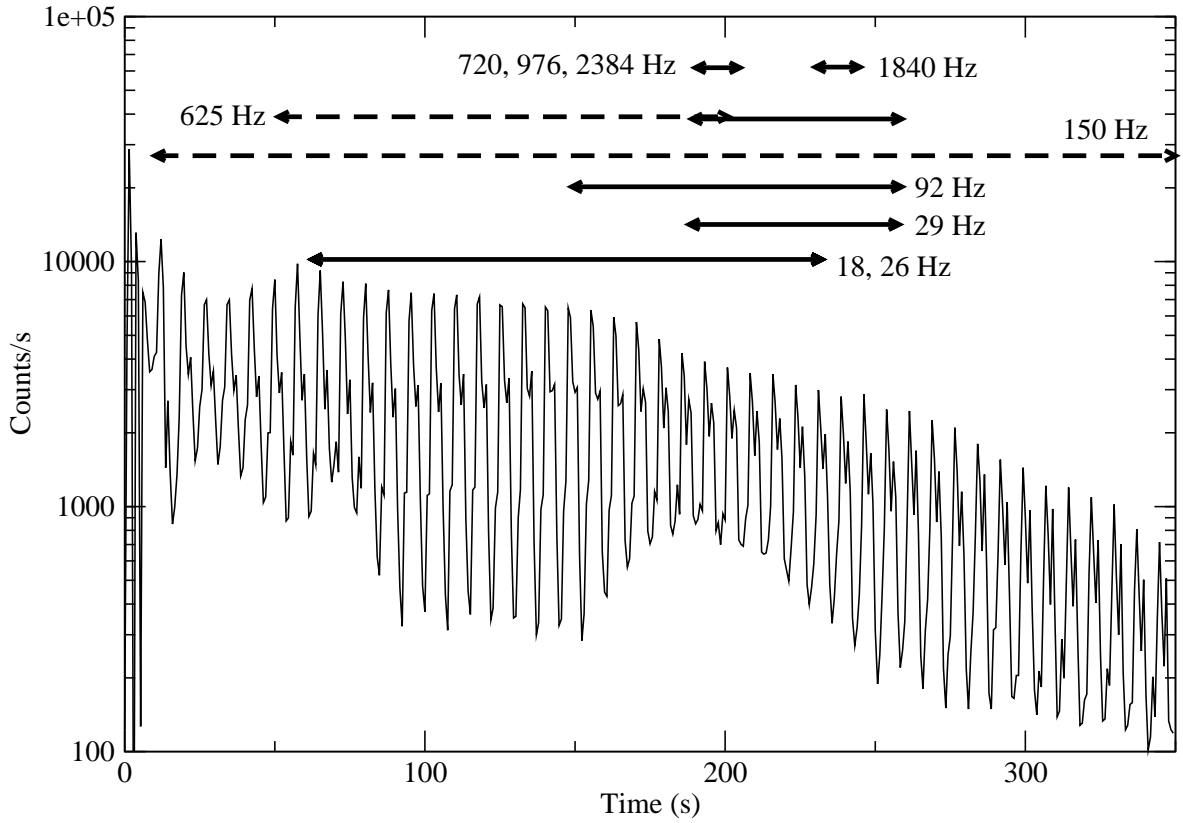}
\end{center}
\caption{The time periods when the different QPOs are detectable in
either the RXTE or RHESSI datasets. Solid lines indicate QPOs that are
detected in the Peak 2/Interpulse region.  Dashed lines indicate QPOs
that are detected primarily during Peak 1. The difference in
properties between the two $\approx$ 625 Hz QPOs, the earlier one
detected by RHESSI and the later one by RXTE, is clear.  }
\label{fig10}
\end{figure}

\pagebreak

\begin{deluxetable}{ccccccc}
\tabletypesize{\scriptsize} \tablecaption{Summary of properties for
the most significant QPOs detected in the tail of the SGR 1806-20
giant flare.  The frequency is the centroid frequency from a
Lorentzian fit; the quoted width is the associated full width at half
maximum.  All amplitudes are the values computed from rotational phase
dependent power spectra. The duration is given with respect to the
main flare at time zero. For the phase, P2/I indicates that the QPO is
concentrated in the Peak 2/Interpulse region; P1 indicates that it is
detected mainly during Peak 1.  The nominal energy band in which the
QPOs are seen is $< 100$ keV except where noted.  Notes: [1]
\citet{wat06}; [2] This paper; [3] \citet{isr05}; [4] The amplitude in
\citet{wat06} was incorrectly reported as 10.0 $\pm$ 0.3 \%; [5] The
amplitude differs from that reported in the previous two rows because
we focus on a different time period [6] Observed in the nominal
100-200 keV band \label{qposum}} \tablewidth{0pt}
\tablehead{\colhead{Frequency (Hz)} & \colhead{Width (Hz)} &
\colhead{RMS amplitude (\%)} & \colhead{Duration (s)} &
\colhead{Phase} & \colhead{Satellite} & \colhead{Notes} } \startdata

 17.9 $\pm$ 0.1 & 1.9 $\pm$ 0.2 & 4.0 $\pm$ 0.3 & 60-230 & P2/I & RHESSI &
    [1] \\[6pt]

 25.7 $\pm$ 0.1 & 3.0 $\pm$ 0.2 & 5.0 $\pm$ 0.3 & 60-230 & P2/I & RHESSI &
    [1] \\[6pt]

 29.0 $\pm$ 0.4 & 4.1 $\pm$ 0.5 & 20.5$\pm$ 3.0 & 190-260 & P2/I & RXTE &
    [2] \\[6pt] 

 92.5 $\pm$ 0.2 & $1.7^{+ 0.7}_{-0.4}$ & 10.7 $\pm$ 1.2 & 150-260 &
    P2/I & 
    RXTE & [3]\\[6pt]

 92.7 $\pm$ 0.1 & 2.3 $\pm$ 0.2 & 10.3 $\pm$ 0.8 & 150-260 & P2/I & RHESSI
    & [1,4]  \\[6pt]

 92.9 $\pm$ 0.2 & 2.4 $\pm$ 0.3 & 19.2 $\pm$ 2.0 & 190-260 & P2/I &
    RXTE & [2,5]\\[6pt]

 150.3 $\pm$ 1.6 & 17 $\pm$ 5 & 6.8 $\pm$ 1.3 & 10-350 & P1 & RXTE &
     [2] \\ [6pt]

 626.46 $\pm$ 0.02 & 0.8 $\pm$ 0.1 & 20 $\pm$ 3 & 50-200 & P1 & RHESSI
 & [1,6] \\ [6pt]

 625.5 $\pm$ 0.2 & 1.8 $\pm$ 0.4 & 8.5 $\pm$ 1.8 & 190-260 & P2/I &
     RXTE & [2] \\ [6pt]

1837 $\pm$ 0.8 & 4.7 $\pm$ 1.2 & 18.0 $\pm$ 3.6 & 230-245 & P2/I & RXTE &
[2] \\[6pt]

  \\ \enddata
  \end{deluxetable}

\end{document}